\documentclass[preprint,12pt]{elsarticle}
\usepackage{graphicx}
\usepackage{dcolumn}
\usepackage{bm}
\usepackage{amsmath}
\usepackage{amssymb}
\usepackage{algorithmic}
\usepackage{graphicx}
\usepackage{textcomp}
\usepackage{xcolor}
\usepackage{quantikz}
\usepackage{adjustbox}
\usepackage{subcaption}
\usepackage{url}

\journal{Physics Letters A}

\begin{document}

\begin{frontmatter}
	
	
	\title{An Enhanced Hybrid HHL Algorithm \tnoteref{label1}}
	\tnotetext[label1]{The first two authors acknowledge the financial support of the IonQ QLAB Global Users Program. The second author also acknowledges the financial support from an IBM Global University Program Academic Award. We thank Romina Yalovetzky, Dylan Herman, and Nikitas Stamatopoulos for fruitful discussions regarding variations of Hybrid HHL.}
	\author{Jack Morgan}
	\ead{morganj@business.unc.edu}
	\affiliation{organization={UNC Chapel Hill},
		country={USA}}
	
	\author{Eric Ghysels}
	\ead{eghysels@unc.edu}
	\affiliation{organization={UNC Chapel Hill},
		country={USA}}
	
	\author{Hamed Mohammadbagherpoor}
	\ead{hamed.mohammadbagherpoor@ibm.com}
	\affiliation{organization={IBM Quantum},
		country={USA}
	}
	
	\begin{abstract}
		We present a classical enhancement to improve the accuracy of the Hybrid variant (Hybrid HHL) of the quantum algorithm for solving linear systems of equations proposed by Harrow, Hassidim, and Lloyd (HHL). We achieve this by using higher precision quantum estimates of the eigenvalues relevant to the linear system, and a new classical step to guide the eigenvalue inversion part of Hybrid HHL. We show that eigenvalue estimates with just two extra bits of precision result in tighter error bounds for our Enhanced Hybrid HHL compared to HHL. Our enhancement reduces the error of Hybrid HHL by an average of 57 percent on an ideal quantum processor for a representative sample of 2x2 systems. On IBM Torino and IonQ Aria-1 hardware, we see that the error of Enhanced Hybrid HHL is on average 13 percent and 20 percent (respectively) less than that of HHL for the same set of systems.
	\end{abstract}
	
	\begin{highlights}
		\item We introduce a classical enhancement to improve the accuracy of Hybrid HHL.
		\item We analytically derive error bounds to prove that Hybrid HHL with our enhancement is more accurate than canonical HHL for a certain class of problems.
		\item We show that Enhanced Hybrid HHL solves a representative sample of problems with 50\% less error than canonical HHL.
	\end{highlights}
	
	\begin{keyword}
		HHL \sep Hybrid Algorithms \sep NISQ Algorithms \sep Linear Algebra
	\end{keyword}
	
\end{frontmatter}

\section{Introduction} \label{sec:intro}
The eponymous HHL algorithm for solving systems of linear equations proposed by Harrow, Hassidim, and Lloyd \cite{Harrow_2009} provides an exponential speedup over classical algorithms, with respect to the size of the linear system $N$. The canonical, or original, HHL algorithm has proven challenging to implement on current hardware for systems with $N$ $>$ 2, see \cite{Aaronson_2005}. A Hybrid HHL algorithm was introduced by \cite{Lee_2019} which can solve a linear system with a shorter circuit depth, if the eigenvalues relevant to the system are estimated separately by an isolated Quantum Phase Estimation (QPE) circuit, which we call the preprocessing step. Moreover, \cite{yalovetzky2023hybrid} shows that the result of Hybrid HHL can be more accurate with higher precision estimates of the eigenvalues. Their proposed method requires that the experimenter manually select eigenvalues based on the results of QPE, and does not address which decision the experimenter should make in a variety of edge cases.

\begin{table}[b]
	\centering
	\begin{tabular}{|c|c|c|}
		\hline
		Algorithm & HHL Circuit & Preprocessing \\
		\hline
		Canonical & $k$ & NA \\
		\hline
		Hybrid & $k$ & $k$ \\
		\hline
		Enhanced & $k$ & $l$ \\
		\hline
	\end{tabular}
	\caption{The number of evaluation qubits used for QPE in the main circuit, and preprocessing circuit for each variant of HHL we study.}
	\label{tab:hhlnot}
\end{table}  

We propose a fully automated end-to-end procedure to perform a more accurate Hybrid HHL algorithm when the number of bits of the preprocessing eigenvalue estimates $l$ is greater than the number of qubits used for QPE within the HHL circuit $k$. We clarify the delineation between Canonical HHL, Hybrid HHL, and our algorithm which we call Enhanced Hybrid HHL in Table \ref{tab:hhlnot}. For each eigenvalue $\lambda_j$, HHL uses a $k$ bit approximation which takes one of $2^k-1$ different values to calculate a rotation angle $\theta (\tilde{\lambda}_k)$ of a Pauli-Y gate within the circuit. Most of the error in the HHL algorithm arises from the difference between $\tilde{\lambda}_k$ and $\lambda_j$. In the edge case where $\lambda_j = \tilde{\lambda}_k,$ the algorithm can be perfectly accurate. In other cases the error tends to scale with the difference between $\lambda_k$ and its estimate. We specify the subscript notation we will be using throughout this paper in Table \ref{tab:eignot}. Reference \cite{yalovetzky2023hybrid} uses $l=4$ bit preprocessing to implement a $k=3$ Hybrid HHL circuit, and ultimately shows a 20\% increase in ideal fidelity with this technique. Their process relies on the experimenter to manually select relevant eigenvalue estimates based on quantum hardware and simulator results. The authors of \cite{yalovetzky2023hybrid} interpret their results as showing the importance of a classical procedure like the one we propose.

\begin{table}
	\centering
	\begin{tabular}{|c|c|c|c|}
		\hline 
		Value & Number of bits & Number of values & Index\\
		\hline
		$\lambda_j$ & exact value & $N$ & $j$\\
		\hline
		$\tilde{\lambda}_k$ & $k$ & $2^k$ & $k$\\
		\hline 
		$\tilde{\lambda}_l$ & $l$ & $2^l$ & $l$\\
		\hline
	\end{tabular}
	\caption{The index and bit precision of $\lambda$ depending on the subscript. The eigenvalue estimates $\tilde{\lambda}_k$ and $\tilde{\lambda}_l$ can take one of discrete number of values shown in the third column. The subscript $k$ denotes both the number of bits, and the index of the estimate. For example, $\sum_{k=0}^{2^k} \tilde{\lambda}_k$ is the sum of every possible $k$ bit estimates.}
	\label{tab:eignot}
\end{table}  

Our procedure consists of three classical steps which we present in section \ref{sec:enhancement}. First, we approximate the value known in the literature as $\alpha_{k|j},$ which is part of the amplitude of cross product of $\vert\tilde{\lambda}_k \rangle$ and the $j^{th}$ eigenvector of the system. Second, for each $\vert \tilde{\lambda}\rangle$ that is entangled at least partially with an eigenvector, we calculate $\theta (\tilde{\lambda}_0 ... )$ that will minimize the error depending on the application. Finally, we filter which rotations exceed a certain threshold of relevance to maintain the circuit depth advantages of Hybrid HHL.

The complexity scaling of Hybrid HHL depends on how the eigenvalue estimates are obtained, which jeopardizes the exponential speedup of the complete algorithm. Classically calculating the eigenvalue estimates completely eliminates the quantum speedup, thus we use a QPE preprocessing circuit to obtain the eigenvalue estimates. The classical processing of the result of QPE is linear with $l$. When $l>\log_2(N)$ this preprocessing eliminates the theoretical speedup. In certain applications the number of eigenvectors that constitute a substantial proportion of the solution vector $m$ is constant. Therefore, while the exponential speedup is lost in theory, the Hybrid Algorithm maintains the benefit in practice. We solve example problems that fit this criterion in Section \ref{sec:implementation}. 

While Hybrid HHL is primarily useful as a proof of concept on noisy processors, Enhanced Hybrid HHL offers a potential use case even after a quantum advantage is achieved with Canonical HHL. When $l<\log_2N$, Hybrid HHL maintains the exponential speedup over classical methods with respect to $N.$ The size and precision of current implementations of HHL are limited by the circuit complexity which scales with $\mathcal{O}\left(\text{log}(N)\kappa^2\right)$ where $N$ is the dimension of the linear system and $\kappa$ is its condition number. On a low noise processor of the distant future, the precision of HHL estimates may be limited by the finite number of qubits in the processor. Reference \cite{yalovetzky2023hybrid} demonstrate how to obtain arbitrarily precise eigenvalue estimates with a fixed number of qubits using a semi-classical inverse Quantum Fourier Transform \cite{Griffiths_1996} to modify the QPE algorithm. Therefore, our procedure to maximize the accuracy of the algorithm with a constant number of qubits remains relevant as quantum processors continue to develop.

The structure of the paper is as follows.  In Section \ref{sec:enhancement} we introduce the enhancement procedure. In Section \ref{sec:error} we compare the error bounds of Enhanced Hybrid HHL to that of Canonical HHL. Finally, Section \ref{sec:implementation} compares the performance of the three variants at inverting a selection of quantum linear system problems on \textit{ibm\_torino} \cite{IBM} and IonQ Aria-1 \cite{IonQ} quantum processors. 

\section{HHL and Hybrid HHL}  \label{sec:hhl}
The Canonical HHL algorithm for linear algebra consists of a single quantum circuit, which we call an HHL circuit. The Hybrid HHL algorithm proposed by \cite{Lee_2019}, is a three step algorithm that uses a preprocessing quantum circuit and a classical computing step (steps 1 and 2), to ultimately execute a shortened HHL circuit (step 3). 

All variations of the HHL algorithm solve a Quantum Linear System Problem (QLSP). A QLSP is defined by an $N$ dimensional linear system $A \vec{x} = \vec{b}$ where $A$ and $\vec{b}$ are known but $\vec{x}$ is not. The matrix $A$ is generally assumed to be Hermitian with a spectral norm of 1. If the spectral norm is not 1, then the matrix can be scaled in a manner explored is Section \ref{sec:Hybrid HHL}. If $A$ is not Hermitian, then the linear system can be reframed as a $2N$ system using the substitutions
{\small
	\begin{equation*}
		\label{eq:HermitianC}
		A_{2N} = 
		\begin{pmatrix} 0 & A_N \\ \overline{A}_N & 0 \end{pmatrix} \qquad b_{2N} = \begin{pmatrix} b_N \\ 0 \end{pmatrix} \qquad x_{2N} = \begin{pmatrix}
			0 \\ \bar{\nu}_{N}  \end{pmatrix}.
\end{equation*}}
To gain an intuitive sense of the workings of HHL, we begin by rewriting the matrix $A$ as a linear combination of the outer products of its eigenvectors, i.e. $\ket{u_i} \bra{u_i}$ weighted by their eigenvalues, yielding:
\begin{equation*}
	A = \sum_{j=0}^{N-1} \lambda_j \ket{u_j}\bra{u_j} 
\end{equation*}
and therefore
\begin{equation*}
	A^{-1} = \sum_{j=0}^{N-1} \lambda_j^{-1} \ket{u_j}\bra{u_j}.
\end{equation*}
If we write $\ket{b}$ in terms of the eigenvectors of $A,$ namely
$\ket{b}$ = $\sum_{j=0}^{N-1} \beta_j \ket{u_j},$
we find the ideal solution can be represented by
\begin{equation}
	\label{eq:sum}
	\ket{x} = \sum_{j=0}^{N-1} \lambda_j^{-1} \beta_j \ket{u_j}.
\end{equation}
In this section we provide a more detailed walk through of how the HHL and Hybrid circuits create an estimate of equation (\ref{eq:sum}). 
\subsection{HHL Circuit \label{sec:HHL}}
The HHL circuit has four main components, namely
state preparation, quantum phase estimation (QPE),
eigenvalue inversion, and inverse quantum phase estimation
(IQPE). These components prepare the approximate solution state $\ket{\tilde{x}}$. We also touch on the ``fifth component'', which is how the approximate solution is utilized or observed. This step will vary depending on the application for which HHL is being used, however this step is critical to maintaining the exponential speedup. In the remainder of this section we closely mirror \cite{Zaman_2023} who provide a step-by-step description of the HHL algorithm as well as numerical examples. 
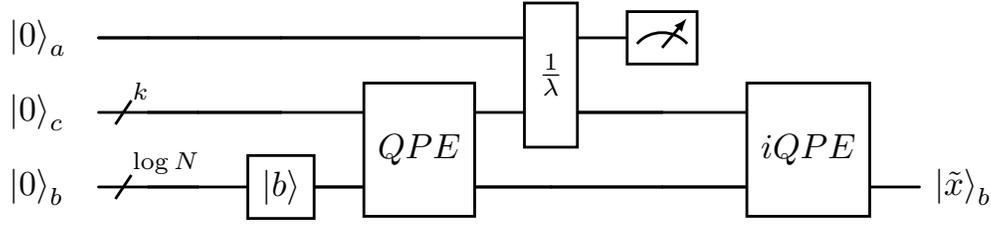
\begin{figure*}
	\centering
	\begin{tikzpicture}
		\node[scale=1.5, midway](circ1){
			\begin{adjustbox}{max size={0.5\textwidth}{0.5\textheight}}
				\begin{quantikz}[column sep={0.5cm}, row sep={0.75cm,between origins}]
					\lstick[label style={text width=0.75cm, align=left}]{$\ket{0}_a$} & \qw & \qw & \qw & \qw & \gate[2]{\frac{1}{\lambda}} & \meter{} & \ctrl[vertical wire=c]{1} \setwiretype{c}  \\
					\lstick[label style={text width=0.75cm, align=left}]{$\ket{0}_{c}$} & \qwbundle{k} & \qw & \qw & \gate[2]{QPE} &  & \qw & \gate[2]{IQPE} \\
					\lstick[label style={text width=0.75cm, align=left}]{$\ket{0}_{b}$} & \qwbundle{\log{N}} & \qw & \gate{\ket{b}} & \qw & \qw & \qw & \qw & \rstick[label style={text width=0.75cm, align=left}]{$\ket{\tilde{x}}_{b}$}\\
				\end{quantikz}
			\end{adjustbox}
		};
	\end{tikzpicture}
	\caption{High level diagram of an HHL circuit with quantum conditional logic controlling the IQPE step. The eigenvalue inversion ($1/\lambda$) sub-circuit is what distinguishes the HHL circuit used between the three variants of the algorithm.}
	\label{fig:hhlcirc}
\end{figure*}

The HHL circuit depicted in Fig.~\ref{fig:hhlcirc} consists of three registers. The ancilla register $a$ consists of one qubit. The clock register $c$ consists of $k$ qubits that are used to evaluate quantum phase estimation. The $b$ register is made of $\log_2(N)$ qubits, which will ultimately hold $\vert \tilde{x} \rangle$. 
We can write the starting state of all three registers as
\begin{equation}
	\label{eq:HHLstep0}
	\ket{\Psi_0} = \ket{0 \ldots 0}_b \ket{0 \ldots 0}_c \ket{0}_a = \ket{0}^{\otimes \log(N)}_b \ket{0}^{\otimes k}_c \ket{0}_a.
\end{equation}
We will show how each step of HHL alters this state to ultimately produce the desired solution. We will present the case when all values of $\lambda_j$ are perfectly approximated by one value of $\tilde{\lambda}_k$ in this exercise for the sake of clarity. Sections \ref{sec:enhancement} and \ref{sec:error} will explore the complications in a more realistic example.

\noindent {\bf State preparation:} The first step is a state preparation circuit which prepares the vector $\ket{b}$ from our problem definition in the $b$ register. The resulting state is
\begin{equation}
	\label{eq:HHLstep1}
	\ket{\Psi_1} = \ket{b}_b \ket{0 \ldots 0}_c \ket{0}_a.
\end{equation}
The construction of the state preparation circuit is an interesting field of study, however it is not the focus of our work. We use the standard \textit{Qiskit} state preparation procedure \cite{qiskit2024}, \cite{Shende_2005}.

\noindent {\bf Quantum phase estimation (QPE):} QPE, which among other things is an eigenvalue estimation algorithm, has three components.
Namely the superposition of the clock qubits through
Hadamard gates, controlled unitary application, and Inverse Quantum Fourier Transform (IQFT). In the first step of QPE, Hadamard gates are applied to the clock qubits to create the state
\begin{equation}
	\label{eq:HHLstep2}
	\ket{\Psi_{2a}} = I^{\otimes b} \otimes H^{\otimes k} \otimes I \otimes \ket{\Psi_1} = \ket{b}_b \frac{1}{\sqrt{2^k}}(\ket{0}_c + \ket{1}_c)^{\otimes n} \ket{0}_a.
\end{equation}
Next we do the controlled unitary application. The controlled gates are applied to $\ket b$ with the clock qubits as the controlling qubits. We apply the unitary operators $U^{2^{r}}$ = $\exp{[iAt_0 2^{r}/T]}$ for $r$ = $0$, $\dots,$ $k-1$ to the $b$ register, controlled by the $r^{th}$ qubit of the $c$ register. The constants $t_0$ and $T$ will be chosen later. When the control clock qubit is $\ket 0,$ $\ket b$ will not be affected but when the clock bit is $\ket 1,$ the unitary rotation operator will be applied. As a result, we see the state
\begin{align}
	\ket{\Psi_{2b}} =&\ket{b}_b \otimes \left(\frac{1}{2^{\frac{k}{2}}}(\ket 0_c + e^{[2\pi i t_0 2^{k-1}/ T]} \ket 1_c)\right) \otimes \left(\frac{1}{2^{\frac{k}{2}}}(\ket 0_c + e^{[2\pi i t_0 2^{2k-2}/ T]} \ket 1_c)\right) \notag
	\otimes \nonumber\\ 
	& \ldots \otimes \left(\frac{1}{2^{\frac{k}{2}}}(\ket 0_c + e^{[2\pi i t_0 2^{0} / T]} \ket 1_c)\right) \otimes \ket{0}_a
\end{align}
to which we apply the IQFT (see equations (15) through (20) in \cite{Zaman_2023}) we get
\begin{align}
	\label{eq:HHLstep4}
	\ket{\Psi_{2c}}  &= \ket b_b \frac{1}{2^{\frac{k}{2}}} \sum_{k=0}^{2^k - 1} e^{[2\pi i t_0 k]} \ket{\tilde{\lambda}_{k}}_c \ket{0}_a \nonumber\\
	&= \sum_{j=0}^{N-1} \beta_j \ket{u_j}_b \ket{\tilde{\lambda}_{k}}_c \ket{0}_a. 
\end{align}
Here we see that the state $\vert \tilde{\lambda}_{k} \rangle$ is the binary bit string of integer $\tilde{\lambda}_{k} t_0 / 2 \pi$.
\noindent {\bf Eigenvalue Inversion:} The eigenvalue inversion sub-circuit entangles the clock register state $\ket{k}$ with the ancilla qubit state\footnote{The states $\ket{0}$ and $\ket{1}$ could be switched in this definition without consequence, however we always use $\ket{1}$ as the $\ket{\text{Good}}$ state to simplify notation.} 
\begin{equation}
	\label{eq:h}
	\vert h(\tilde{\lambda}_k) \rangle  = \sqrt{1- h(\tilde{\lambda}_k)^2} \ket{0} + h(\tilde{\lambda}_k) \ket{1} 
\end{equation}
where
\begin{equation*}
	h(\tilde{\lambda}_k) = \begin{cases} 
		0 & \tilde{\lambda}_k \leq C \\
		\frac{C}{\tilde{\lambda}_k} & \tilde{\lambda}_k \geq C
	\end{cases}.
\end{equation*}
The constant $C$ is an underestimate of the smallest (by absolute value) relevant eigenvalue of the QLSP chosen by the experimenter, based on the limited information that is known about the problem in advance. The circuit implementing this step is a distinguishing factor between the variations of HHL, which we explore in further detail in Section \ref{sec:Hybrid HHL}. Adjoining $\vert h(\tilde{\lambda}_k) \rangle$ to $\ket{\Psi_{2c}}$ results in 
\begin{align}
	\label{eq:HHLstep5}
	\vert \Psi_3 \rangle  &= \sum_{j=0}^{N-1} \beta_j \ket{u_j}_b \ket{\tilde{\lambda}_k}_c \vert h(\tilde{\lambda}_k)\rangle_a \nonumber \\
	&= \sum_{j=0}^{N-1} \beta_j \ket{u_j}_b \ket{\tilde{\lambda}_k}_c \left(\text{Garbage}\ket{0} + \frac{C}{\tilde{\lambda}_k} \ket{1} \right).
\end{align}

When executing HHL on a processor with mid-circuit measurement capabilities, we measure the ancilla qubit to complete eigenvalue inversion. If the state does not collapse to $\ket{1}_a$ then we discard and execute the circuit again from step 1. On processors without this capability, we complete the circuit and measure the ancilla along with our final measurement in step 5. For steps 4 and 5, we will only consider the results where the ancilla qubit is measured in the $\ket{1}_a$ state.

\noindent {\bf Inverse quantum phase estimation
	(IQPE):} After successful eigenvalue inversion, we need to disentangle the clock register with our solution state. We accomplish this with the inverse of step 2. The result is the state 
\begin{equation}
	\label{eq:HHLstep7}
	\ket{\Psi_4}  =  \sum_{j=0}^{N - 1} \frac{\beta_j C}{\tilde{\lambda}_k} \ket{u_i}_b \ket{0}_c^{\times k} \ket{1}_a = \ket{\tilde{x}}_b \ket{0}_c^{\otimes k} \ket{1}_a
\end{equation}
and therefore the approximate solution is stored in the $b$-register. This walk through pertains to the ideal case where $\lambda_j = \tilde{\lambda}_k$ and thus the state created by HHL will be an exact solution to the QLSP. This assumption will not be true for most implementations, which is why we distinguish the state prepared by HHL $\ket{\tilde{x}}$ and the true solution $\ket{x}$.

\noindent {\bf Measurement:} The final step of HHL is dependent on the applicaiton for which the circuit is being used. Directly measuring $\ket{\tilde{x}}$ poses no exponential speedup compared to classically computing $\vec{x}$, thus defeating the purpose of using HHL to begin with. The algorithm should be used either as a portion of a larger quantum algorithm that requires the state $\ket{x}$, or for applications where the desired end result is the expectation value of an operator observing $\ket{x}$, see \cite{ghysels2023quantum} for examples.

\begin{figure}
	\begin{tikzpicture}
		\node[scale=1, midway](circ1){
			\begin{adjustbox}{width=\columnwidth}
				\begin{quantikz}[column sep={0.5cm}, row sep={0.75cm,between origins}]
					\lstick[label style={text width=0.75cm, align=left}]{$\ket{\tilde{x}}_{b}$} & \qwbundle{\log{N}} & \qw & \qw & \qw & \swap{1} & \qw & \\
					\lstick[label style={text width=0.75cm, align=left}]{$\ket{0}_{st}$} & \qwbundle{\log{N}} & \qw & \qw & \gate{\ket{x}} & \swap{0} & \qw & \qw  \\
					\lstick[label style={text width=0.75cm, align=left}]{$\ket{0}_{st\ a}$} & \qw & \qw & \qw & \gate{H} & \ctrl{-1} & \gate{H} & \meter{} \\
				\end{quantikz}
			\end{adjustbox}
		};
	\end{tikzpicture}
	\caption{Full circuit diagram of the SWAP test used to measure the accuracy of the estimate $\ket{\tilde{x}}$ produced by an HHL circuit. The gate $\ket{x}$ is the default state preparation algorithm provided by \textit{qiskit} using the classically computed solution state $\ket{x}.$}
	\label{fig:stcirc}
\end{figure}
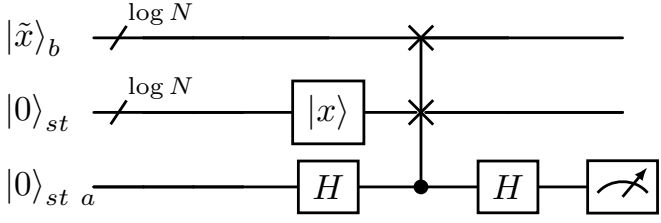

We are interested in the error of HHL and it's variants. The error defined by $||\ket{x} - \ket{\tilde{x}}||$ can be extrated from the fidelity $\braket{\tilde{x}}{x}$ using 
\begin{equation*}
	||\ket{x} - \ket{\tilde{x}}|| = \sqrt{2 \left(1-\mathcal{R}\{\braket{\tilde{x}}{x}\}\right)}.
\end{equation*}
In our simulator trials we obtain $\braket{\tilde{x}}{x}$ by measuring the expectation value $\bra{\tilde{x}}P_x\ket{\tilde{x}}$ where $P_x$ is the classically computed projection operator onto $\ket{x}$. In our main tests, we use a controlled-SWAP test \cite{Beausoleil_2008} like the one displayed in Fig.~\ref{fig:stcirc} to measure the inner product between these two states. In \ref{app:iqm} we omit the SWAP test and directly measure the $b$ register to obtain our estimate of $\ket{\tilde{x}}$, which we then use to calculate the error. To perform the controlled SWAP test, we add $\log_2N$ qubits to our circuit in the so called SWAP test ($st$) register, and a SWAP test ancilla ($st\ a$) qubit. The SWAP test circuit follows 5 steps:
\begin{itemize}
	\item Prepare the classically computed $\ket{x}$ state in the $st$ register.
	\item Apply a Hadamard gate to $st\ a.$
	\item Use the $st\ a$ qubit to control a SWAP gate between the $st$ and $b$ registers.
	\item Apply a second Hadamard gate to $st\ a.$
	\item Measure $st\ a$.
\end{itemize}

We calculate $\mathcal{R}\{\braket{\tilde{x}}{x}\}$ using the measurement probabilities of the circuit shown in Fig.~\ref{fig:stcirc} with the formula
\begin{equation*}
	\mathcal{R}\{\braket{\tilde{x}}{x}\} = \sqrt{2\frac{P(\ket{10})}{P(\ket{10})+P(\ket{11})-1}},
\end{equation*}
which ultimately leads us to the error defined by
\begin{equation*}
	||\ket{\tilde{x}} - \ket{x}|| = \sqrt{2(1-\mathcal{R}\{\braket{\tilde{x}}{x}\})}.
\end{equation*}

\subsection{Hybrid HHL Algorithm}
\label{sec:Hybrid HHL}
The Hybrid HHL algorithm proposed by \cite{Lee_2019} differs from Canonical HHL in two key ways. First, a preprocessing quantumm circuit is used to estimate $\lambda_j$ and $\beta_j$ for the eigenvectors of $A$. Second, the ancilla qubit rotation portion of HHL is replaced with a sub-circuit that only performs the rotation on the $m$ most relevant eigenvalues where $m \leq N$. Relevance is the amplitude of said eigenvector in the solution $\ket{x}$, which we approximate with $\beta_k / \lambda_k$. The combination of these measures allow for a reduction in the depth of the HHL circuit.

In our hardware implementations, we follow \cite{Lee_2019} who use the QPE algorithm. This algorithm is identical to measuring the clock register after the second step of HHL. The output of this circuit yields the $\beta_k^2$ for $2^k$ different bitstrings which correspond to the different values of $\tilde{\lambda}_k$. Reference \cite{yalovetzky2023hybrid} outlines an exciting alternative to QPE which promises arbitrary bit precision estimates with a fixed number of qubits. Their approach uses the semi-classical IQFT \cite{Griffiths_1996} which replaces the two qubit gates in the standard IQFT with classically controlled single qubit gates. As a result, the eigenvalue estimate is stored on a singular clock qubit which is measured and reused instead of a the clock register with $k$ qubits. With this technique a processor capable of running dynamic quantum circuits \cite{Corcoles_2021} like the semi-classical IQFT can produce higher precision eigenvalue estimates without increasing the number of qubits in the circuit. 

Along with this semi-classical preprocessing method, \cite{yalovetzky2023hybrid} proposes an iterative process during preprocessing to find the optimal choice of $t_0$ that results in the largest $\lambda_j$ being estimated by the largest possible $\tilde{\lambda}_k$. This processes is a combination of the two steps they call ``Algorithm 1'' and ``Algorithm 2''. The goal of Algorithm 1 is to find the optimal $t$ such that the greatest eigenvalue $\lambda_{max}$ corresponds to the highest possible $\tilde{\lambda}_k$ without overflowing. The first iteration in Algorithm 1 requires an excessively small value of $t_0$ to ensure that all eigenvalue ``estimates'' round to 0. In subsequent iterations, $t_0$ is increased in a manner that converges to the optimal choice. Algorithm 2 checks for the edge case where the initial value of $t_0$ was not sufficiently small, and all eigenvalues happened correspond to an overflow estimate that is a multiple of $2^k$ and thus represented by $\ket{0}^{\otimes k}$. This is verified by running QPE with an even smaller value of $t_0$ to confirm that all eigenvalues still round to 0. Reference \cite{yalovetzky2023hybrid} provides a more detailed breakdown of both algorithms. We confirm that this method yields more accurate results on a noiseless simulator in section \ref{sec:implementation}, however the cost is too prohibitive for our hardware results.

Now we turn our attention to the main Hybrid HHL circuit, which follows the first two steps in the Canonical HHL circuit. The advantage of the hybrid approach comes in at the eigenvalue inversion step. In Canonical HHL this step can be implemented by either a V-Chain \cite{Barenco_1995}, Gray Code \cite{Mottonen_2004}, or Uniformly Controlled Rotation (UCR) \cite{Barenco_1995} sub-circuits. Reference \cite{yalovetzky2023hybrid} provides a detailed comparison of these methods. In keeping with the potential use case for Enhanced Hybrid HHL, we want to compare Hybrid HHL inversion to the Canonical HHL inversion option that is best suited for a situation where the number of qubits available on an extremely high fidelity processor is limiting $N$ and $k$. Thus we rule out the V-Chain method because it is the only method that requires additional ancilla qubits to implement. For $k \geq 6,$ the Gray Code requires more CNOT gates to implement than a Uniformly Controlled Rotation. For this reason, we exclusively compare Hybrid HHL inversion to UCR in our analysis. 

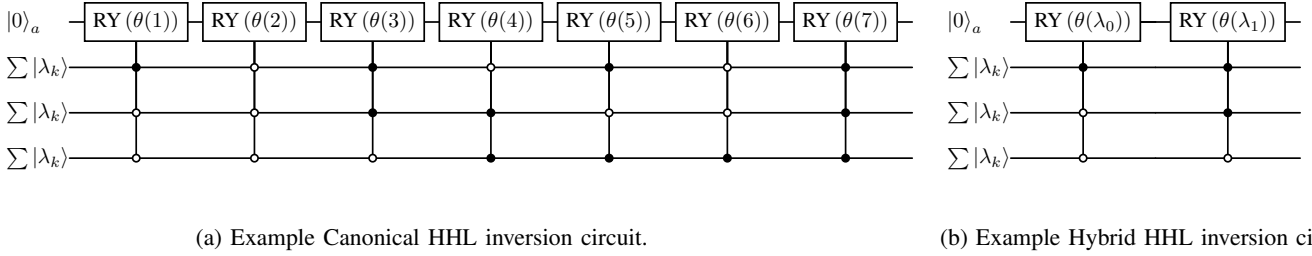
\begin{figure*}
	\begin{subfigure}[t]{0.62\textwidth}
		\begin{adjustbox}{height=0.05\textheight}
			\begin{quantikz}[column sep={0.25cm}, row sep={0.75cm,between origins}]
				\lstick[label style={text width=0.9cm, align=left}]{$\ket{0}_a$} & \gate{\text{RY}\left(\theta({1})\right)} & \gate{\text{RY}\left(\theta(2)\right)} & \gate{\text{RY}\left(\theta(3)\right)} & \gate{\text{RY}\left(\theta(4)\right)} & \gate{\text{RY}\left(\theta(5)\right)} & \gate{\text{RY}\left(\theta(6)\right)} & \gate{\text{RY}\left(\theta(7)\right)} & \\
				\lstick[label style={text width=0.9cm, align=left}]{$\sum\ket{\lambda_k}$} & \ctrl{-1} & \octrl{-1} & \ctrl{-1} & \octrl{-1} & \ctrl{-1} & \octrl{-1} & \ctrl{-1} & \\
				\lstick[label style={text width=0.9cm, align=left}]{$\sum\ket{\lambda_k}$} & \octrl{-2} & \octrl{-2} & \ctrl{-2} & \ctrl{-2} & \octrl{-2} & \octrl{-2} & \ctrl{-2} & \\
				\lstick[label style={text width=0.9cm, align=left}]{$\sum\ket{\lambda_k}$} & \octrl{-3} & \octrl{-3} & \octrl{-3} & \ctrl{-3} & \ctrl{-3} & \ctrl{-3} & \ctrl{-3} &  \\
			\end{quantikz}	
		\end{adjustbox}
		\caption{Example Canonical HHL inversion circuit.} \label{fig:cann_inversion}
	\end{subfigure}%
	\hspace*{\fill}   
	\begin{subfigure}[t]{0.31\textwidth}
		\begin{adjustbox}{height=0.05\textheight}
			\begin{quantikz}[column sep={0.25cm}, row sep={0.75cm,between origins}]
				\lstick[label style={text width=0.9cm, align=left}]{$\ket{0}_a$} & \gate{\text{RY}\left(\theta({\lambda_0})\right)} & \qw & \gate{\text{RY}\left(\theta({\lambda_1})\right)} &  \\
				\lstick[label style={text width=0.9cm, align=left}]{$\sum\ket{\lambda_k}$} & \ctrl{-1} &  & \ctrl{-1} & \\
				\lstick[label style={text width=0.9cm, align=left}]{$\sum\ket{\lambda_k}$} & \octrl{-2} & & \ctrl{-2} & \\
				\lstick[label style={text width=0.9cm, align=left}]{$\sum\ket{\lambda_k}$} & \octrl{-3} & & \octrl{-3} &  \\
			\end{quantikz}
		\end{adjustbox}
		\caption{Example Hybrid HHL inversion circuit.} \label{fig:hybrid_inversion}
	\end{subfigure}
	\caption{Comparison of the inversion circuits in the Canonical ~\ref{fig:cann_inversion} and an example Hybrid ~\ref{fig:hybrid_inversion} HHL circuits with $3$ clock qubits. In this example implementation, preprocessing only found two relevant rotations.} \label{fig:inversion_cirucits}
\end{figure*}

A UCR sub-circuit is made up of $k$ clock qubits and one ancilla qubit. Each state $\vert\tilde{\lambda}_k \rangle \leq C$ controls a Pauli-Y rotation (RY) applied to the ancilla qubit. The angle of said rotations is calculated with
\begin{equation*}
	\theta(k) =  2 \arcsin{\left(h(\tilde{\lambda}_k)\right)}
\end{equation*}
using the definition in equation (\ref{eq:h}). In the most general case this results in $2^k-1$ rotations. Fig.~\ref{fig:cann_inversion} shows an example of an uniformly controlled rotation for $k=3$.

The Hybrid eigenvalue inversion circuit, like the one seen in Fig.~\ref{fig:hybrid_inversion}, is similar to the UCRY however it only contains the rotations that exceed a chosen relevance threshold. Omitting irrelevant rotations reduces the circuit depth and ultimately the hardware noise, however it introduces error into the estimate. Let $\tilde{\lambda}_{k = 0}, \tilde{\lambda}_{k = 1}, \ldots, \tilde{\lambda}_{k = r_k}$ be the relevant eigenvalue estimates, and $\vert \tilde{\Psi} \rangle$ be the state created by the Hybrid HHL circuit which differs from the canonical state starting with the eigenvalue inversion step. Instead of the state shown in equation (\ref{eq:HHLstep5}), the Hybrid eigenvalue inversion circuit results in 
\begin{align*}
	\ket{\tilde{\Psi}_3} = & \sum_{j=0}^{r_k} \beta_j \ket{u_j} \vert \tilde{\lambda}_k \rangle \left(\sqrt{1 - \frac{C^2}{\tilde{\lambda}_k^2}} \ket{0}_a + \frac{C}{\tilde{\lambda}_k} \ket{1}_a\right) + \nonumber\\
	& \sum_{j=r_k}^{N-1} \beta_j \ket{u_j} \vert \tilde{\lambda}_k \rangle \ket{0}_a. 
\end{align*}
After a measurement of the ancilla qubit, the state of the shots that collapse to the $\ket{1}$ state is
\begin{equation*}
	\ket{\tilde{\Psi}_4}  = \sum_{j=0}^{r_k} \beta_i \ket{u_i} \vert \tilde{\lambda}_j \rangle \frac{C}{\tilde{\lambda}_j} \ket{1}_a, 
\end{equation*}
which differs from the state of an ideal Canonical HHL by the omitted rotations
\begin{equation}
	\label{eq:hybrid_inversion_error}
	\ket{\Psi_4} - \vert \tilde{\Psi}_4 \rangle  = \sum_{j=r_k}^{N} \beta_i \ket{u_j} \vert \tilde{\lambda}_j \rangle \frac{C}{\tilde{\lambda}_j} \ket{1}_a. 
\end{equation}
Hybrid HHL is useful when the ideal error to the final solution that results from equation (\ref{eq:hybrid_inversion_error}) is offset by the decrease in the circuit depth and thus number of qubit errors when implemented on a noisy QPU. 

\section{Enhanced Hybrid HHL}
\label{sec:enhancement}
We propose a classical addition to the Hybrid HHL algorithm that mathematically selects the rotation angles and control states for the eigenvalue inversion sub-circuit based on the results of preprocessing. We call the complete algorithm with our addition Enhanced Hybrid HHL.
The output of $l$ bit preprocessing yields the $r_l$ most relevant values of $\tilde{\lambda}_l$, as well as 
\begin{equation*}
	\beta_l = \sum_{j=0}^N \beta_j \ \alpha_{l|j}
\end{equation*}
for each relevant estimate, where the amplitude $0 \leq \alpha_{k|j} \leq 1$ is a function of the distance between $\tilde{\lambda}_k$ and $\lambda_j$. Our enhancement relies on the substitutions
\begin{align}
	\label{eq:substitutions}
	& \lambda_j \approx \tilde{\lambda}_l \nonumber \\
	& \beta_j \approx \beta_l \nonumber \\
	& \alpha_{k|j} \approx \alpha_{k|l}.
\end{align}
In other words, we treat the higher precision eigenvalue estimates as the exact values. With this framework established we introduce our enhancement consisting of three parts:
\begin{itemize}
	\item Determine $\alpha_{k|l}$ for the two adjacent $\tilde{\lambda}_k$ values for each relevant $\tilde{\lambda}_l$.
	\item Choose the rotation angle such that minimizes the approximate error for each control state $\vert \tilde{\lambda}_k \rangle$.
	\item Filter the irrelevant states based on the threshold criteria used to reduce the circuit depth.
\end{itemize}
\noindent {\bf Determine $\mathbf{\alpha_{k|j}}$:} Let's revisit step 2 of HHL, but now we consider the realistic case where $\lambda_j$ is not perfectly estimated. Under these conditions, the state produced by QPE takes the form 
\begin{equation}
	\label{eq:alphasv}
	\ket{\tilde{\Psi}_{2c}} = \sum_{j=0}^{N-1} \sum_{k=0}^{2^k} \beta_j \ \alpha_{k|j} \ket{u_j}_b \ket{\tilde{\lambda}_{k}}_c \ket{0}_a.
\end{equation}
Reference \cite{phillips2024detailed} shows that $\alpha_{k|j}$ can be simplified to 
\begin{equation}
	\label{eq:alpha}
	\alpha_{k|j} = \frac{\sqrt{2}}{T}\sin{\frac{\pi}{2T}}\frac{\left|\cos{\frac{\delta}{2 T}} \cos{\frac{\delta}{2}}\right|}{\left|\sin{\frac{\delta+\pi}{2T}} \sin{\frac{\delta - \pi}{2 T}}\right|},
\end{equation}
where $\delta$ is the difference in phase between $\lambda_j$ and $\tilde{\lambda}_{k}$. The $\delta$ between two adjacent values of $\tilde{\lambda}_{k}$ is $2\pi/t_0$. In the region $0 < \delta <2\pi,$ we approximate equation (\ref{eq:alpha}) with
\begin{equation*}
	\alpha_{k|j} \approx \alpha_{k|l} \approx 1 - \frac{\delta_{k|l}}{2 \pi}. 
\end{equation*} 
The mean squared error of this approximation 0.16. For every $\tilde{\lambda}_{l}$ whose $x_l$ exceeds the relevance threshold from our preprocessing circuit, we calculate $x_l$ $\alpha_{k|l}$ for the two adjacent values of $\vert \tilde{\lambda}_{k} \rangle.$

\noindent {\bf Choose the rotation angles:} Now we calculate the optimal angle for each controlled rotation in the Hybrid eigenvalue inversion circuit. To do this, we define the vector $\vec{v}_k$ that is added to the final estimate by each individual rotation. We use standard vector notation here because these states will not be normalized, as implied by the definition
\begin{align*}
	\ket{\tilde{x}} = \sum_{k=0}^{2^k}  \vec{v}_k.
\end{align*}
We can apply HHL steps 3 and 4 to equation (\ref{eq:alphasv}) to find that in the realistic case
\begin{equation*}
	\ket{\tilde{x}} = \sum_{k=0}^{2^k} \sum_{j=0}^{N-1} \beta_j \ \alpha_{j|k} \ \bar{x}_k \ket{u_j},
\end{equation*}
from which we deduce
\begin{equation}
	\label{eq:rot_mag}
	\vec{v}_k = \sum_{j=0}^{N} \bar{x}_k \ \beta_j \ \alpha_{k|j} \ket{u_j}
\end{equation}
where
\begin{equation}
	\label{eq:barx}
	\bar{x}_k = \frac{\sin(\theta(k))}{2}.
\end{equation} 
Let us also define the hypothetical vector $\vec{w}_k$ in the summation
\begin{align*}
	\ket{x} = \sum_{k=0}^{2^k} \vec{w}_k
\end{align*}
which takes the form
\begin{equation*}
	\vec{w}_k = \sum_{j=0}^{N} \ \beta_j \ \alpha_{k|j} \ \lambda_j^{-1} \ket{u_j}.
\end{equation*}
The best way for us to minimize the total error $||\ket{x} - \ket{\tilde{x}}||$ without complete information about the ideal solution is to find the value of $\bar{x}_k$ that minimizes $||\vec{v}_k - \vec{w}_k||$ for each relevant value of $k$. We derive that the latter magnitude is
\begin{equation*}
	||\vec{v}_k - \vec{w}_k|| = \sum_{j=0}^{N} \left|\alpha_{k|j} \ \beta_{j} \left(\lambda_j^{-1} - \bar{x}_k\right)\right|^2,
\end{equation*}
which we can rewrite using the substitutions in equation (\ref{eq:substitutions}) to see
\begin{equation}
	||\vec{v}_k - \vec{w}_k|| \approx \sum_{l=0}^{r_l} \left|\alpha_{k|l} \ \beta_l \left(\tilde{\lambda}_l^{-1} - \bar{x}_k\right)\right|^2
\end{equation}
The choice of $\bar{x}_k$ that minimizes this function is a weighted average of $\lambda^{-1}_l \ \forall \ l \mid \alpha_{k|l}>0$, with $\beta_l \  \tilde{\alpha}_{k|l}$ as the weights. We then use this this value and equation (\ref{eq:barx}) to determine that the rotation angle is
\begin{equation*}
	\theta_k = \arcsin\left(\frac{2}{r_l} \sum_{l=0}^{r_l} \frac{\alpha_{k|l} \ \beta_l}{\tilde{\lambda}_l}\right).
\end{equation*}

\noindent {\bf Apply threshold filter:} Along with the relevance threshold applied to the measurement results of preprocessing in Hybrid HHL, we apply a second filter to remove rotations caused by a relevant $\tilde{\lambda}_l$ with an amplitude of entanglement with $\vert \tilde{\lambda}_k \rangle$ that is irrelevant. We define relevance in this context as 
\begin{equation*}
	r_{\theta_k} = \sum_{l = 0}^{r_l} \frac{\beta_l \ \tilde{\alpha}_{k|l}} {\tilde{\lambda}_l}.
\end{equation*}
As with Hybrid HHL, threshold that Enhanced Hybrid HHL uses to determine relevance is once again chosen based on the desired circuit depth and accuracy. Further research could determine the optimal threshold based on device noise characteristics.

\section{Error Analysis \label{sec:error}}
Deriving completely automated algorithm allows us to perform analytic error analysis to prove that our enhancement reduces the error bounds of Hybrid HHL. 
Our error analysis closely follows the proof of the no post selection case derived by \cite{Harrow_2009}. For this analysis, we will assume that the relevance threshold is below the lowest nonzero $\beta_j$. Recall that in the perfectly estimated case without mid-circuit measurement takes the form
\begin{equation*}
	\ket{x} = \sum_{j=0}^{N-1} \beta_j \ket{u_j}_b \ket{0}^{\otimes k} \vert h\left(\lambda_j\right) \rangle_a.
\end{equation*}
Starting from the solution, we can recreate the ideal version $\ket{\Psi_3}$ by applying the inverse of IQPE (also known as QPE) to the solution which yields
\begin{equation*}
	\ket{\Psi_3} = \sum_{j=0}^{N-1} \beta_j \sum_k \alpha_{j|k} \ket{k}_c \ket{u_j}_b \vert h\left(\lambda_j\right) \rangle_a.
\end{equation*} 
We will derive the error of the algorithm by comparing the fidelity between this state and the actual state at this portion of the algorithm
\begin{equation*}
	\vert \tilde{\Psi}_3 \rangle = \sum_{j=0}^{N-1} \beta_j \sum_k \alpha_{j|k} \ket{k}_c \ket{u_j}_b \vert h\left(\lambda_j\right) \rangle_a,
\end{equation*}
from which we derive the inner product
\begin{equation*}
	\langle \tilde{\Psi}_3 \vert \Psi_3 \rangle = \sum_{j=0}^{N-1}\left|\beta_j\right|^2 \sum_{k=0}^{2^k} \left|\alpha_{k|j}\right|^2 \langle h(\tilde{\lambda}_k) \vert h(\lambda_j) \rangle.
\end{equation*}
We can treat $j, k$ like random variables whose joint distribution takes the form $\sum_{j=0}^{N-1}\left|\beta_j\right|^2 \sum_k \left|\alpha_{k|j}\right|^2$, and therefore
\begin{equation*}
	\text{Re}\langle \tilde{\Psi}_3 \vert \Psi_3 \rangle = \mathbb{E}_j \mathbb{E}_k \langle h(\tilde{\lambda}_k) \vert h(\lambda_j) \rangle
\end{equation*}
The proof of Lemma 2 from \cite{Harrow_2009} shows that the squared norm of the derivative of $h(\lambda)$ with respect to $\lambda$ is $ \leq \kappa^2$. Using this bound we find
\begin{equation*}
	|| \ket{h(\lambda_1)} - \ket{h(\lambda_2)} || \leq c \kappa \left|\lambda_1 - \lambda_2 \right|,
\end{equation*} 
from which we deduce
\begin{equation*}
	\text{Re}\langle h(\tilde{\lambda})_l \vert h(\tilde{\lambda}_j) \rangle \geq 1 - \frac{c^2 \delta^2 \kappa^2}{2 t_0^2}, 
\end{equation*}
where $c\leq 1$ is a constant, $\delta = t_0 |\lambda_j - \lambda_k|,$ and
\begin{equation*}
	t_0 = \frac{2 \pi (2^k -1)}{\lambda_{k max}}.
\end{equation*} 
Next, we divide the sources of infidelity into two camps; one for the two closest values of $\ket{k}$, and a second for all other control states. In Canonical HHL, the two closest values of $\lambda_k$ correspond to the distance $\delta \leq 2 \pi$. We can apply the tighter bound $\delta \leq 2 \pi / (2^{(l-k)})$ when using $l$ bit estimates because there exists $2^{(l-k)}$ evenly spaced values of $\tilde{\lambda}_l$ between each $\tilde{\lambda}_k$. Therefore, the first source of infidelity for Enhanced HHL is less or equal to  $2 \pi^2 \kappa^2 / t_0^2 (2^{(l-k)}).$

For all other values of $k$, we assume that $\vert \tilde{\lambda}_k \rangle$ is not a relevant control state, and thus $\vert h(\tilde{\lambda}_k) \rangle = \ket{0}.$ The smallest inner product between the ideal inversion function and $\ket{0}$ comes from the minimum eigenvalue $\lambda_{j\ min}$, where we see that
\begin{align}
	\label{eq:second_camp}
	\vert h(\lambda_{j min}) \rangle &= \sqrt{1-\frac{C^2}{\lambda_{j\ min}^2}}\ket{0} + \frac{C}{\lambda_{j\ min}}\ket{1} \nonumber\\
	&= \frac{C}{\lambda_{j\ min}}\ket{1},
\end{align} 
for the ideal choice of $C$ which is $C = \tilde{\lambda}_{j\ min}$. We rewrite this choice of $C$ as
\begin{equation*}
	C = \frac{2 \pi k_{min}}{t_0},
\end{equation*} 
where $k_{min}$ is the integer representation of $\tilde{\lambda}_k$. We assume that $k_{min}=1$, as is true when $\kappa / t_0$ is large. This assumption is fair because as we will see the error scales with $\kappa / t_0$. Since both states are normalized and perpendicular to each other, we rewrite the worst case scenario as
\begin{equation*}
	||\ket{0} - \ket{h(\lambda_{j\ min})}|| \leq \sqrt{2} \frac{C}{\lambda_{j min}} = \sqrt{2} \frac{\frac{2 \pi}{t_0}}{\frac{1}{\kappa}} = \sqrt{2}\frac{2 \pi \kappa}{t_0}.
\end{equation*}
Finally, we use our new bound for $||\ket{0} - \ket{h(\lambda_{j})}||$ and $|\alpha_{k|j}|^2 \leq 64 \pi^2/\delta^4$ to find that the second source of infidelity is 
\begin{equation*} 
	\leq \sum_{k=\frac{\lambda_k t_0}{2 \pi}+1}^{\infty} \frac{64 \pi^2}{\delta^4} \frac{8 \pi^2 \kappa^2}{t_0^2} = \sum_{k=1}^{\infty} 32 \pi^4 \frac{\kappa^2}{t_0^2} \frac{1}{\pi^4 k^4} = \frac{32}{90}\pi^4 \frac{\kappa^2}{t_0^2}
\end{equation*}
where we use $\sum_{k=1}^{\infty} 1/k$ = $\pi^4/90,$ 
to simplify. Combining the two sources we achieve an ultimate infidelity of 
\begin{equation*}
	\text{Re}\langle \tilde{\Psi}_3 \vert \Psi_3\rangle  = 1 - \left(\frac{2}{2^{(l-k)}} + \frac{32 \pi^2}{45}\right)\pi^2 \frac{\kappa^2}{t_0^2}
\end{equation*}
which corresponds to 
\begin{equation*}
	\| \vert \tilde{\Psi}_3 \rangle  - \ket{\Psi_3}\| \leq \sqrt{\frac{1}{\pi^2 2^{(l-k)}} + \frac{16}{45}} 2 \pi^2 \frac{\kappa}{t_0}.
\end{equation*}
We observe that for $l = k+2$, the square root term works out to equal $0.62,$ whereas for HHL without enhancement, $l=k$ the same term equals $0.68$. The original error bounds calculated for HHL is
\begin{equation*}
	\| \vert \tilde{\Psi}_3 \rangle - \ket{\Psi_3} \| \leq 2 \pi^2 \frac{\kappa}{t_0},
\end{equation*}
which would make the enhanced hybrid error bound $38\%$ tighter than the canonical version. Reference \cite{phillips2024detailed} updates the calculations made by \cite{Harrow_2009} and find 
\begin{equation*}
	\| \vert \tilde{\Psi}_3 \rangle - \ket{\Psi_3} \| \leq \sqrt{\frac{20}{3}} c \pi \frac{\kappa}{t_0},
\end{equation*}
where $c \leq \pi / 2$, thus putting the revised canonical error bound between that of Hybrid and Enhanced Hybrid HHL.

\begin{figure}
	\centering
	\begin{adjustbox}{width=0.6\textwidth}
		\includegraphics[scale=1]{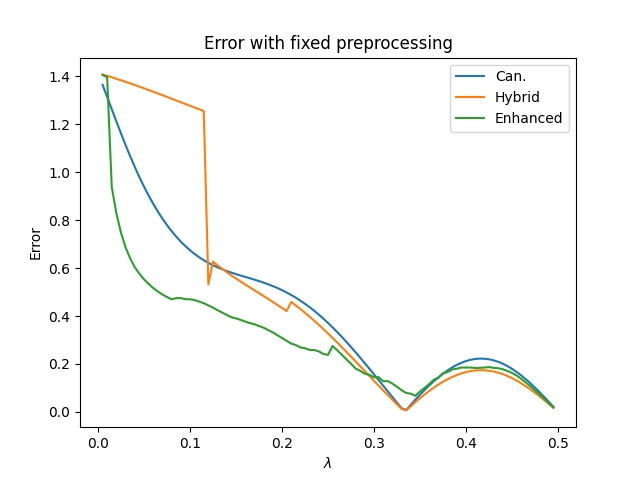}
	\end{adjustbox}
	\caption{Error curves for the three variants solving the linear system problems defined by equation \ref{eq:exampleqlsp} on the noiseless \textit{ibm\_qasm\_simulator}. The system corresponding to $\lambda = 1/3$ bears eigenvalues of $1/3$ and $2/3$ which are perfectly estimated by the $k=3$ clock qubits used for this test.}
	\label{fig:errorcurve} 
\end{figure}

\section{Implementation}\label{sec:implementation}
Errors in the estimate produced by the HHL circuit come from the algorithmic errors, as well as hardware noise. Hybrid and Enhanced Hybrid HHL use shorter eigenvalue inversion sub-circuits, which reduces the impact of hardware noise on the final result. We implement all three algorithms to solve the test quantum linear system problems outlined in this Section. All example problems and code used to implement the algorithms can be found in \cite{Morgan_Enhanced-Hybrid-HHL_2024}. On all tests run on IonQ Aria we use $1024$ shots, while tests run on \textit{ibm\_torino} we use the standard $4000$ shots. In all experiments we use $l=5$ with a preprocessing relevance threshold of $2^{-l}$, and $k=3$ with an enhancement relevance threshold of $2^{-k}.$ We find from our $N=2$ and $N=4$ test QLSPs that Enhanced Hybrid HHL is on average slightly more accurate than the other two variants on today's quantum processors. However, based on the noiseless simulator results and circuit depth calculations we expect this difference to be greater on more accurate processors.

\subsection{N=2 Matrix Implementation}
We begin by following \cite{Lee_2019} in characterizing the error of each algorithm with a set of $N=2$ quantum linear systems defined by
\begin{equation}
	\label{eq:exampleqlsp}
	A_{2N} = 
	\begin{pmatrix} 0.5 & \lambda - 0.5 \\ \lambda - 0.5 & 0.5 \end{pmatrix} \qquad b_{2N} = \begin{pmatrix} 1 \\ 0 \end{pmatrix}
\end{equation}
which results in eigenvalues $\lambda_{j0} = \lambda$ and $\lambda_{j1} = 1-\lambda$ and an equal projection of $b$ onto each eigenvector. We compare the ideal accuracy of all three variants of HHL on the \textit{ibm\_qasm\_simulator} for 99 values of $\lambda$ between $0$. We confirm the findings of \cite{Lee_2019} that the error of the classical and hybrid algorithms is $0$ when $\lambda_{j}$ is perfectly estimated by $\lambda_k$. The sharp spikes in Fig.~\ref{fig:errorcurve} for Hybrid and Enhanced Hybrid curves correspond to cases where estimated eigenvalues pass the preprocessing relevance threshold. Enhanced Hybrid HHL has the lowest error accross the range of eigenvalues that are poorly estimated by $k$ qubits, and a lower average error across the entire range. All algorithms are nearly perfectly inaccurate with $\lambda = 0.01$ because the linear system is poorly conditioned ($\kappa = 99$). The average error shown in Table \ref{tab:idealaverageerror} is consistent with the ordering of our error bounds; Enhanced Hybrid is more accurate than Canonical which is more accurate than Hybrid. 
\begin{table}
	\centering
	\begin{tabular}{|l|c|c|}
		\hline
		Algorithm & Error & Error with Iterative Preprocessing \\
		\hline
		Canonical & 0.43 & 0.43 \\	
		\hline
		Hybrid & 0.51 & 0.49 \\
		\hline
		Enhanced Hybrid & 0.31 & 0.21 \\
		\hline
	\end{tabular}
	\caption{The average ideal error of the HHL implementations in Fig.~\ref{fig:errorcurve}, and the ideal error of the same problems solved with the iterative preprocessing method outlined in Section \ref{sec:Hybrid HHL}. As referenced, the iterative preprocessing procedure scales $t_0$ to the optimal value for the particular QLSP, which results in a lower error. All tests were conducted on the \textit{ibm\_qasm\_simulator}.}
	\label{tab:idealaverageerror}
\end{table}

For our hardware tests, we look at the quantum linear systems defined by equation \ref{eq:exampleqlsp} for values of 
\begin{align*}
	\lambda \in \{3/24,&\ 4/24,\ 5/24,\ 6/24,\ 7/24, \\
	&8/24,\ 9/24,\ 10/24,\ 11/24\}.
\end{align*} 
We choose these values of $\lambda$ because they are evenly spaced throughout the range in which equation 
(\ref{eq:exampleqlsp}) is well conditioned ($1.1 < \kappa < 6.4.$). Based on Fig.~\ref{fig:errorcurve}, the benefit of our enhancement is even greater when $A_{2N}$ is poorly conditioned, however we focused on problems that could realistically be solved by $k=3$ HHL.

We show the average error of these tests in Table \ref{tab:N2test} and the detailed error of each individual test in the examples folder of \cite{Morgan_Enhanced-Hybrid-HHL_2024}. Across both processors, we find that Enhanced Hybrid HHL is more accurate than the other two variants. We observe that the difference between the performance of Enhanced Hybrid HHL and the other variants is greater on an ideal processor than on actual hardware. This is due to noise in the preprocessing circuit, which results in an inaccurate enhancement calculation. When we run the tests with simulated preprocessing results, we see a greater reduction in the error with our enhancement. Using simulated preprocessing results and the IonQ Aria-1 processor we recorded an average error of 0.54, which is the most accurate average of any of the tests.
\begin{table}
	\centering
	\begin{subtable}[t]{0.45\textwidth}
		\centering
		\begin{tabular}{|c|c|c|}
			\hline
			Algorithm & \textit{ibm\_torino} & Aria-1 \\
			\hline
			Canonical & 0.68 & 0.75 \\
			\hline
			Hybrid & 0.68 & 0.60 \\
			\hline
			Enhanced & 0.59 & 0.59 \\
			\hline
		\end{tabular}
		\caption{Measured Error}
		\label{tab:N2error}
	\end{subtable}
	\hfill 
	\begin{subtable}[t]{0.45\textwidth}
		\centering
		\begin{tabular}{|c|c|c|}
			\hline
			Algorithm & \textit{ibm\_torino} & Aria-1 \\
			\hline
			Canonical & 1219 & 288 \\
			\hline
			Hybrid & 541 & 157 \\
			\hline
			Enhanced & 674 & 155 \\
			\hline
		\end{tabular}
		\caption{HHL Circuit Depth}
		\label{tab:N2depth}
	\end{subtable}
	\caption{The average error (Table \ref{tab:N2error}) and circuit depths (Table \ref{tab:N2error}) of each HHL algorithm solving our $N=2$ test QLSPs on \textit{ibm\_torino} and IonQ Aria-1. The final row shows the test result when we use classically simulated preprocessing results.} 
	\label{tab:N2test}
\end{table}

\subsection{N=4 Matrix Implementation}
The largest matrices that have been inverted with HHL to date are reported in \cite{yalovetzky2023hybrid} and \cite{saevarsson_2022}. We explore how our enhancement can be used to reduce the error of these cutting edge implementations. We searched for a linear system our enhancement has the greatest improvement to the error of the final estimate. To inform our search, we consider what is known about the relationship between the eigenvalue estimates and the error in the estimated state. As shown by \cite{Lee_2019}, and highlighted in Section \ref{sec:error}, all variants have zero error when the relevant eigenvalues are perfectly estimated by $k$ bits. When $\lambda_j$ perfectly bisects $\tilde{\lambda}_k$ and $\tilde{\lambda}_{k+1},$ and thus is the average of those two states, the net effect of the two adjacent Canonical HHL rotations calculated with $\tilde{\lambda}_k$ and $\tilde{\lambda}_{k+1}$ is close to that of the enhanced rotations. Of course, when $\lambda_j$ is close to $\tilde{\lambda}_k$, the effect of a singular hybrid rotation calculated with $\tilde{\lambda}_k$ is also close to that of the enhanced rotations. Therefore the ideal eigenvalues $\lambda_j$ to showcase our enhancement are 
\begin{equation*}
	\tilde{\lambda}_k \leq \lambda_j \leq \frac{\tilde{\lambda}_k + \tilde{\lambda}_{k+1}}{2}
\end{equation*}
This range corresponds to $\frac{\pi}{2} \leq \delta \leq \pi.$ We cannot derive a clean formula for the exact location thanks to equation (\ref{eq:alpha}), thus we perform simulator tests using an $N=4$ matrices with many eigenvalues spanning the chosen range and select the results that show the most promise. An error in the absolute value of all relevant $\tilde{\lambda}_k$ that maintains the relative size of $\tilde{\lambda}_k^{-1}$ does not induce an error because the solution state $\ket{x}$ is by normalized by measurement. In other words, we want the $\tilde{\lambda}_k < \lambda_j$ for the $\tilde{\lambda}_k$ closest to the smallest $\lambda_j$, and we want the $\tilde{\lambda}_k > \lambda_j$ for the $\tilde{\lambda}_k$ closest to the largest $\lambda_j$. Therefore, we searched over quantum linear system problems with one eigenvalue that is $\delta$ greater than $\lambda_{k=0}$, and another that is $\delta$ less than $\lambda_{k=1}$ for the range $\frac{\pi}{2} \leq \delta \leq \pi.$

Let us consider a clock register of $k=3$ bits using two's compliment code to account for positive and negative eigenvalues. In this register, $\tilde{\lambda}_k$ takes values in $\{-1, -2/3, -1/3, 1/3, 2/3, 1\}$ where we ignore the $\ket{000}$ state because it is poorly conditioned, and the $\ket{100}$ due to overflow. We arbitrarily chose eigenvalues that would be estimated by $\tilde{\lambda}_{k=0} = 1/3$ and $\tilde{\lambda}_{k=1} = -1$. This search leads us to the Hermitian matrix $A_{test}$ with the eigenvalues
\[ \lambda_j \in \{-21/24, -20/24, 5/24, 6/24\}.
\]
We constucted $A_{test}$ by randomly generating eigenvectors, and using the formula
\begin{equation}
	\label{eq:testmat}
	A_{test} = 
	\begin{pmatrix} 
		\vec{u}_0 \\ 
		\vdots \\
		\vec{u}_3 
	\end{pmatrix}
	\begin{pmatrix} 
		\lambda_{j=0} & 0 & 0 \\ 
		0 & \ddots & 0 \\
		0 & 0 & \lambda_{j=3} 
	\end{pmatrix}
	\begin{pmatrix} 
		\vec{u}_0^T & \dots & \vec{u}_3^T \\ 
	\end{pmatrix}
\end{equation}
to generate our test matrix. We construct $\vec{b}_{test\ \lambda_{j=0}\ \lambda_{j=1}}$ to be an equal superposition of the two eigenvectors corresponding to our choices of $\lambda_{j=0}$ and $\lambda_{j=1}$ for a given test. The complete list of test problems are available in \cite{Morgan_Enhanced-Hybrid-HHL_2024}, and the results of this test in Table \ref{tab:N4test}. The accuracy benefit of our enhancement is less relevant on Aria-1 than it would be on an ideal processor. This is due to noise in the preprocessing circuit results, which will ultimately affect the classical calculation. We still observe that the theoretical error is reduced by 20\% relative to the hybrid and canonical algorithms, which gives promise that the enhancement will be more important as less noisy processors are developed. 
\begin{table}[h]
	\centering
	\begin{subtable}[t]{0.45\textwidth}
		\centering
		\begin{tabular}{|c|c|c|}
			\hline
			Algorithm & \textit{ibm\_torino} & Aria-1 \\
			\hline
			Canonical & 1.08 & 1.06 \\
			\hline
			Hybrid & 1.13 & 1.03 \\
			\hline
			Enhanced & 1.08 & 1.1 \\
			\hline
		\end{tabular}
		\caption{Measured Error}
	\end{subtable}
	\hfill 
	\begin{subtable}[t]{0.45\textwidth}
		\centering
		\begin{tabular}{|c|c|c|}
			\hline
			Algorithm & \textit{ibm\_torino} & Aria-1 \\
			\hline
			Canonical & 2802 & 794 \\
			\hline
			Hybrid & 2352 & 675 \\
			\hline
			Enhanced & 2538 & 745 \\
			\hline
		\end{tabular}
		\caption{HHL Circuit Depth}
	\end{subtable}
	\caption{The average error and circuit depths of each HHL algorithm solving our $N=4$ test QLSPs on \textit{ibm\_torino} and IonQ Aria-1. On a noiseless simulator, Canonical HHL and Hybrid HHL both average an error of $0.30$ whereas Enhanced Hybrid HHL averages $0.24$ for these matrices. The final row shows our average error when we use the IonQ simulator to simulate the preprocessing circuit without noise.} 
	\label{tab:N4test}
\end{table}
To the best of our knowledge, the most accurate $N=4$ matrix inverted with HHL on a QPU was performed by \cite{yalovetzky2023hybrid} on the Honeywell H1 processor. Their implementation used 3 clock qubits and 4 bit eigenvalue estimates, and the authors to manually chose relevant eigenvalues based on hardware and simulator preprocessing results. The resulting state from this implementation had an error of 1.04. Across the 12 tests we performed, the only result with an error $<$ 1 was an enhanced hybrid hhl test with an error of 0.99 (fidelity of 0.51), which stands as the only $N=4$ implementation to cross this accuracy milestone. When we follow \cite{yalovetzky2023hybrid} and use simulated preprocessing results we see an even greater increase in accuracy. Using simulated preprocessing results, the average error over the four problems we tested was $0.96$, with the most accurate individual test resulting having an error of $0.88$. Our findings confirm that accurately solving $N=4$ QLSPs still proves to be a difficult task for near term hardware, however Enhanced Hybrid HHL is at the cutting edge in this domain.

\section{Discussion}
We introduce a classical enhancement to the existing Hybrid HHL algorithm that creates a more accurate estimate $\ket{\tilde{x}}$, and requires a shorter circuit depth than Canonical HHL. Our enhancement builds on the work of \cite{yalovetzky2023hybrid} who first demonstrated that $l$ bit eigenvalues can be used to create a more accurate $k$ bit Hybrid HHL cirucit. We provide an automated workflow, and analytic error analysis to prove the reduction in error bounds. We show that with $k+2$ bit eigenvalue estimates, the bounded error of Enhanced Hybrid HHL is 38\% less than the originally derived canonical bounds, while maintaining the same dependence on $\kappa$ and $t_0$. Our enhancement reduces the average error of Hybrid HHL by an average of 13\% and 20\% when solving a representative sample of $N=2$ quantum linear system on \textit{ibm\_torino} and IonQ Aria-1 respectively. We also perform what is to our knowledge the largest matrix inversion with an error $<$ 1 to date using entirely quantum hardware results. This technique bridges the gap between noisy quantum hardware and seamless quantum applications by performing the reduced depth Hybrid HHL circuit without manual preprocessing input from the experimenter.

\appendix
\section{$N=2$ implementation without SWAP Tests }
\label{app:iqm}
As shown in Section \ref{sec:implementation} the limited connectivity of \textit{ibm\_torino} results in a substantially greater circuit depth than the all-to-all connectivity trapped ion processors like IonQ Aria-1. IQM recently unveiled a IQM Deneb, superconducting processor with a unique star-shaped topology. This processor has six qubits and a novel hardware component called the central resonator which allows each qubit to indirectly control a cz gate on any other qubit. This is done with the novel move operation which will excite the resonator if the control qubit is excited. The resonator can then be used to control a cz gate on any other qubit, to effectively recreate all-to-all connectivity with one extra two-qubit operation. We are interested in how IQM Deneb compares to the other superconducting processor we study for this application. IQM Deneb has 6 qubits, and thus cannot run an HHL circuit with the SWAP test that we used in Section \ref{sec:implementation}. Instead, we directly measure $\ket{\tilde{x}}$ and compare the measurement probabilities with the probabilities of $\ket{x}$. We then perform a similar test on the other two processors to provide a fair benchmark. This method does not provide any exponential speedup, and results in a greater sampling noise than the SWAP test results. In this appendix, we analyze the fidelity and circuit depth results from this test. 

 We show the results of this test across the three processors in Table \ref{tab:iqm}. The results from each experiment are averaged over nine problem instances, with 3000 shots per instance. We see a smaller error on \textit{ibm\_torino} in this test compared to Table \ref{tab:N2test} on account of the lack potential qubit errors introduced by the SWAP test circuit. On IQM Deneb, we do not see a drop in error as a result of our enhancement. This aberration from the trends we see on the other two processors could be the result of sampling error, which is greater in this test. Further research could repeat this experiment with a larger sample size or a classical emulator of IQM Deneb to verify that this trend is a result of the hardware characteristics. Assuming for now that this trend would persist with an unlimited sample size, we speculate that single qubit gate error rate may be the distinct characteristic of IQM Dened that is responsible for this trend. The single qubit gates on IQM Deneb have higher error rates the single qubit gates on \textit{ibm\_torino} \cite{IQM} \cite{IBM}. While the topology and lower two-qubit gate error rates result in less overall error of the HHL circuit, this single qubit error rate means that the difference between $\theta_{l}$ and $\theta_{k}$ will be less significant to the circuit result. From these tests we conclude that IQM Deneb is not well suited to benefit from our enhancement.

\begin{table}[h]
	\centering
	\begin{subtable}[t]{0.45\textwidth}
		\centering
		\begin{tabular}{|c|c|c|}
			\hline
			Algorithm & \textit{ibm\_torino} & Deneb \\
			\hline
			Canonical & 0.62 & 0.58 \\
			\hline
			Hybrid & 0.77 & 0.59 \\
			\hline
			Enhanced & 0.59 & 0.59 \\
			\hline
		\end{tabular}
		\caption{Measured Error}
		\label{tab:iqm_error}
	\end{subtable}
	\hfill 
	\begin{subtable}[t]{0.45\textwidth}
		\centering
		\begin{tabular}{|c|c|c|}
			\hline
			Algorithm & \textit{ibm\_torino} & Deneb \\
			\hline
			Canonical & 725 & 458 \\
			\hline
			Hybrid & 338 & 224 \\
			\hline
			Enhanced & 414 & 273 \\
			\hline
		\end{tabular}
		\caption{HHL Circuit Depth}
		\label{tab:iqm_depth}
	\end{subtable}
	\caption{Results of our HHL tests without using SWAP test on \textit{ibm\_torino} and IQM Deneb. The measured error in Table \ref{tab:iqm_error} was calculated based on the measurement probabilities after 3000 shots. For these test used the standard transpiler offered by each organization. The depths of the resulting circuits are in Table \ref{tab:iqm_depth}.} 
	\label{tab:iqm}
\end{table}

\bibliography{apssamp}

	\end{document}